# PERFORMANCE EVALUATION OF THE EFFECT OF NOISE POWER JAMMER ON THE MOBILE BLUETOOTH NETWORK

Adil H. M. Aldlawie[1] and Ghassan A. QasMarrogy[2]

[1,2]Communication and Computer Engineering Department, Cihan University, Erbil, Iraq

## ABSTRACT

*Network jammers can uncover the reliability of the wireless networks, where the position of the jammer node permit the network to cope with jamming leveraging varieties of defence strategies. Wireless Bluetooth network is a short range network, where the Bluetooth transceiver is operating in the 2.4 GHz ISM (industrial, scientific and medical) radio band, and a special type of frequency hopping pattern was included with the specification of the Bluetooth technology to offer the necessity requirement for the matching Radio Frequency RF channels. This paper calculates and evaluates the effect of noise power jammer and follower jammer on the Personal Area Network (PAN) Bluetooth network. The results prove that the effect on the Bluetooth network depends on many factors such as power density, distance and immunity technique, which is used in many Bluetooth networks.*

## 1. INTRODUCTION

Due to the limited number of unauthorized frequency bands, and the growing extensiveness demand on the wireless network technologies, will cause a swarming environment full of radio signals that leads to unintentional interference of radio band shared the same spectrum. Many devices are using the Bluetooth technology such as smartphones, laptops, ipads, and smart watches to communicate with each other, where a Bluetooth is a technique to transfer information by using a short range of RF signals between them. In this paper a PAN Bluetooth network (mobile equipment with a free handset device) will be evaluated and simulated to test the effect of noise powered jamming with different instantaneous bandwidth of jamming, that operates in the 2.4 GHz ISM band, over 79 MHz bandwidth, that bandwidth is dived to 79 channels, each channel has 1 Mbps capacity as shown in table 1. The main objective is to measure the level of the signal jamming that reaches the PAN, and comparing it with a certain level that effects the network. The method of frequency hopping is used across a 79 MHz frequency range to evade the interference between the communicated nodes in the equivalent band, the time for transmission is separated among 625µs slots, where each slot uses a new hop frequency technique. Many types of Bluetooth jammers where used to stop the data transmission between the nodes inside the network, such as follower jammer and noise jammer [1].

Many research studies have been done to evaluate the effect of jammers performance on the wireless Bluetooth network by using different type of simulators, where in [2] The author uses a different type of service denial attack called distributed jammer network (DJN) made of a great number of tiny radio, low-power jammers, the research shows that DJN can cause a phase transition in the performance of the wireless targeted network. The author in [3] calculates the performance of Bluetooth and IEEE 802.11b systems affected by the mutual interference, by developing a simulation framework based on detailed MAC and PHY models for modelling the interference. The research in [4] shows a Wireless Denial of Service (WDoS) attack can be done by using off the shelf device, by repeatedly transmitting an RF signal to jam any access to the network, by using any malicious node along with interfere with the receiver node, also different method was proposed to detect the existence of jammers nodes.

TABLE 1. Specification of Bluetooth network

| Items | Specification |
|---|---|
| Power o/p | 1 mw |
| B.W. | 1 MHz |
| Processing gain | 19 dB |
| Receiver sensitivity | - 70 dBm |
| Equipment distance | ( 0.30 to 1.2 ) meters |

The rest of the paper is organized as follows: Section 2 describes the noise jammer concept. Section 3, shows the types of jamming used. Section 4 demonstrate the effect calculation for noise jammer, and section 5 shows the conclusion.

## 2. NOISE JAMMER CONCEPT

With little effort the developing of radio signal technologies has permitted challengers to construct purposeful jammers to interrupt network communication between the network nodes. Regardless of unintentional interference or malicious jamming. Radio jamming technique is an electromagnetic waves transmitted to the purpose of interrupting the transmission of communication by reducing the SNR (signal to noise ratio). Sometimes jamming take a place when the sender node sends on a full of activity channel without testing that the frequency is in use or not, or when the network nodes unintentionally release a signal, this called unintentional jamming [5].

Therefore, to interrupt the transmission in the wireless network, purposeful wireless jammers is used, where the wireless sender is tuned to the same frequency and the same type of modulation as the opponents' receiving device, and with enough amount of power it can override any signal at the receiver across the targeted network. Wireless jammers has many types, such as recorded sounds, spark, random keyed modulated CW, warbler, tone, random pulse, random noise, stepped tones, pulse, rotary, sweep through and gulls, all the mentioned types can be categorized into two main groups, obvious and subtle [5].

Noise jammer system depend on a noise signal which is generated by noise source, and that signal will be limited by certain bandwidth by a filter, and convert to desired radio frequency RF, then amplified to desire power by an amplifier, then the signal will be transmitted by an antenna. Fig 1 demonstrate the elements that connected together to generate noise jammer.
Where the Fc is the frequency of the carrier and BW is the bandwidth. The filter bandwidth must be equal half of the desired RF bandwidth, while the radio frequency (F) will be chosen within the Bluetooth operation band and the mode of jamming.

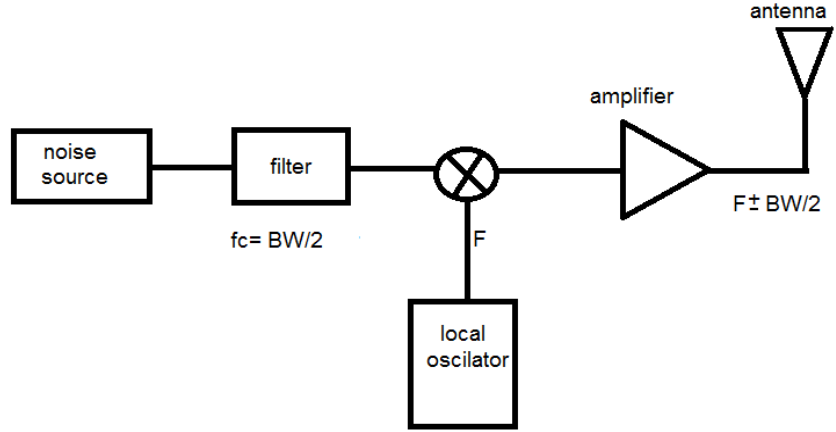

Fig 1. Noise Jammer System.

## 3. TYPES OF JAMMING USED

One of the most commonly wireless jamming technique used is the Barrage noise jamming, in this jammer the Bandwidth BW is more than the BW of the instantaneous signal, and the frequency of the noise jamming signal is expanded to reject the usage of a multiple communication or radar frequencies, to negate successfully the transmission of the data [6], therefore by using this technique, one jammer can simultaneously jam different radar frequencies, but still it has the essential problem were the wider spread of the jamming signal is, the less jamming power available per radar. The other most commonly wireless jamming technique type used is the sweep jammer, where the BW of the jammer can take a part or sub part of the total BW of the wireless transmitted signal, and the full power of the jammer is shifted from one frequency to another. While this has the advantage of being able to jam multiple frequencies in quick succession, it does not affect them all at the same time, and thus it limits the effectiveness of this type of jamming [7]. Therefore, for PAN Bluetooth network, Barrage noise jamming uses all the 79 MHz band, with high power usage, while sweep jamming uses a sub bandwidth like 5 MHz and that will be swept over 79 MHz, The multi sub band jammer operates with 20 MHz band width to cover all the bands and to increase the power density of the jammer signal [8], the time for sweep over the band must be selected carefully according to Adaptive Frequency Hopping (AFH) which will cancel the bad channel with the higher bit error rate number or effected by noise.

## 4. EFFECT CALCULATION FOR NOISE JAMMER

The effect of jamming signals on the Bluetooth network must be calculated for several scenarios, where each scenario has special conditions such as jammer power, bandwidth of the jamming signal, distance between the jammer and the network elements, distances between the elements in network, signal to noise jammer and so on. Therefore three network scenarios where simulated using matlab simulator, the scenarios jammer specification was considered as shown in table 2, 3, and 4:

TABLE 2. Jammer specification scenarios

| Items | 1st scenario | 2nd scenario | 3rd scenario |
|---|---|---|---|
| Power output | 1 , 2 , 5 watt | 1 , 2 , 5 watt | 1 , 2 , 5 watt |
| Bandwidth | 79 MHz barrage jammer | 20 MHz barrage jammer | 5 Mhz. sweep jammer |
| Jammer-network Distance | 10 meters | 10 meters | 10 meters |
| The jamming to signal ratio | > 3 dB | > 3 dB | > 3 dB |
| Power output | 1 , 2 , 5 watt | 1 , 2 , 5 watt | 1 , 2 , 5 watt |

Equation (1) shows how the path loss was calculated

$$PL\ (dB) = 32.4 + 20\ Log\ (D) + 20\ Log\ (F) \quad (1)$$

Where (F) is the carrier frequency 2.440 GHz and (D) is the distance in (meter).

According to equation (1), the effected path loss for the PAN by jamming signal is shown in fig 2. Where the attenuation is increased due to the distance between the jammer and the Bluetooth equipment.

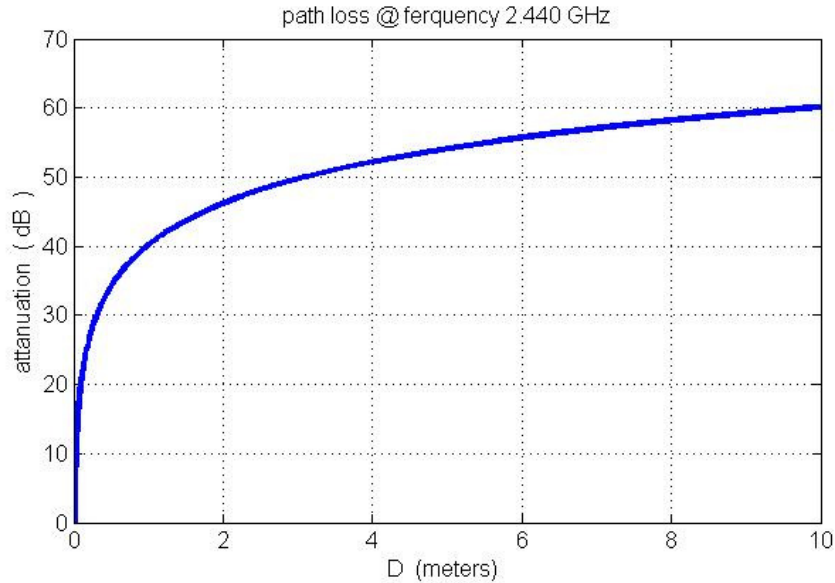

Fig 2. The Path Loss for Jamming Signal to Network Nodes at F=2440 MHz

The minimum jamming signal that reaches the network equipment must be greater than (sensitivity level + processing gain) so that the jamming signal will be defined as the equation (2) with adding the 3db condition for jamming to signal ratio, because 3db is the signal minimum level that must be covered.

$$J > S+PG + 3 \quad (2)$$

J is jamming signal, S the sensitivity of the network equipment, and PG is the processing gain. By substitution all factors in equations (3-4)

$$J = - 70dbm +19\ dB +3 \quad (3)$$
$$J \geq - 48\ dBm \quad (4)$$

Where -70dbm is the sensitivity of the Bluetooth receiver, and 19 dB is the hopping technique processing gain. Therefore, the jamming signal must be greater than the value (-48) dBm to effect the network.

In case of the 1st scenario the effect will not appeared on the network, because the distance is equal to 6m, which give us high attenuation similar to 117dB when is compared with the transmitted power. In case of the 2nd scenario which has 20MHz, the processing gain is reduced with time, due to AFH that reduced the interference between the Bluetooth frequencies, by holding all the channels that effected by this signal and not use it [9], and for the 3rd scenario which has 5 MHz, by using Sweep jammer technique that increases the probability of intercepting the jamming signal with the frequency hopping signal, therefore it increases the number of deleted channels by AFH, as shown in figure 3.

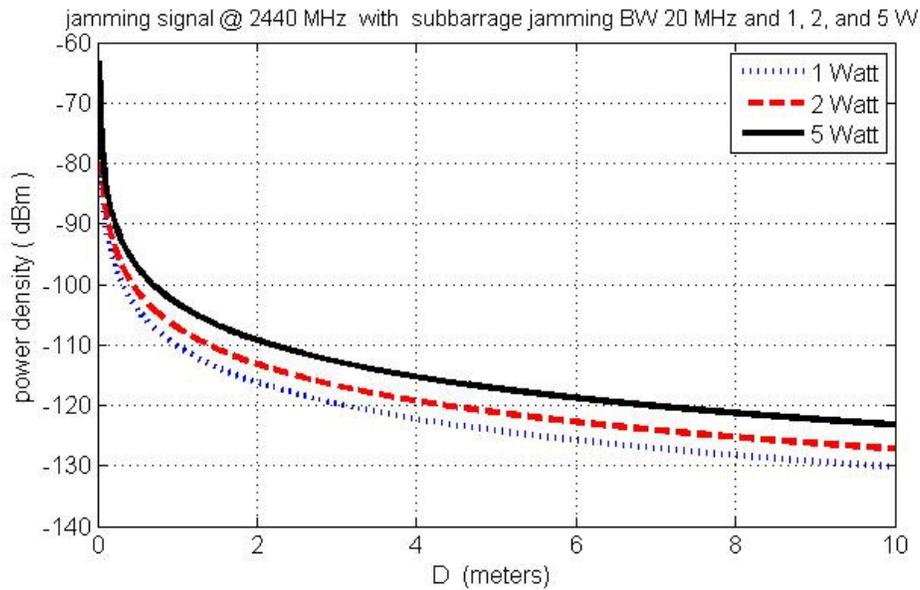

Fig 3. Jamming Signal Levels through the Distance for Different Jamming Power

The Bit Error Rate BER will be increased when the channel is effected by the jamming signal, then that channel will be eliminated from the channels list, and the processing gain will be reduced with time [10], therefore the channel will be effected by the jammer. Figure 4 shows the processing gain reduced with time due to AFH by using sweep jammer [9], and it's happened when the channels effected by noise jammer.

International Journal of Computer Networks & Communications (IJCNC) Vol.6, No.5, September 2014

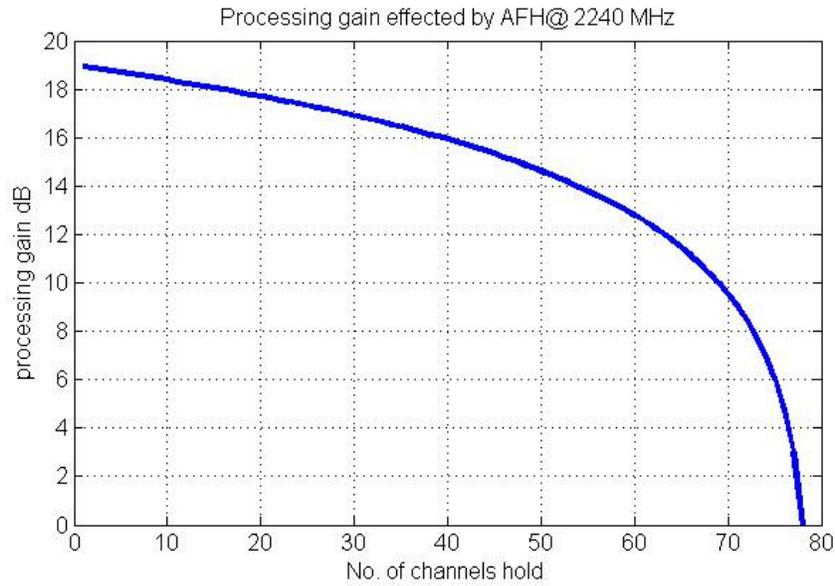

Fig 4. Processing Gain Effected By Sweep Noise with Time

Figure 5 shows the sweep jamming technique, where the signal increases because of AFH technique, which is cancelled the channels that have high BER.

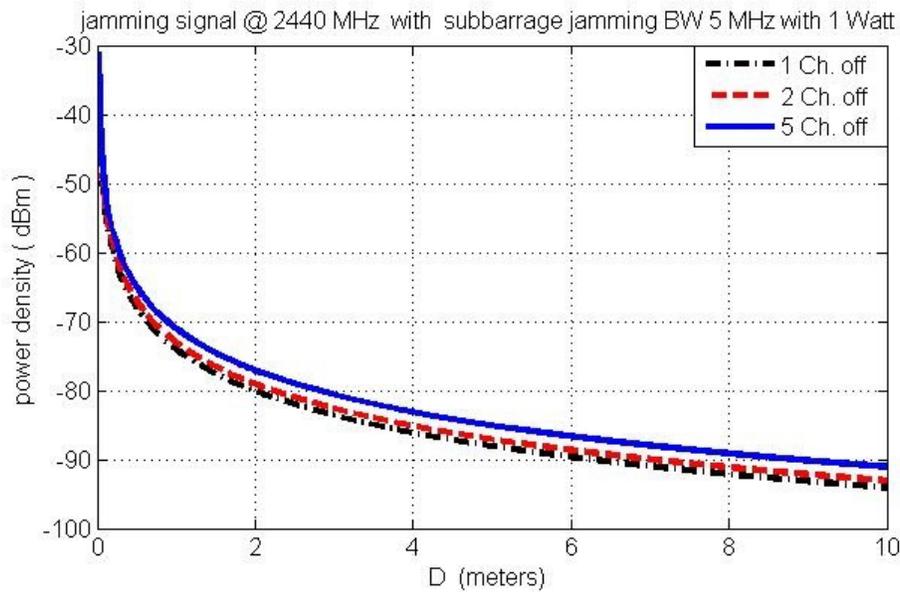

Fig 5. Jamming Signal Level Increase with Time Duty to AFH

Finally by using the power jamming transmitter 2 watt and 5 watt as shown in figures 6 and 7, there is no difference between power levels in all the scenarios when it compared with the level jamming requirement (-48dbm) from 6m distance.

172

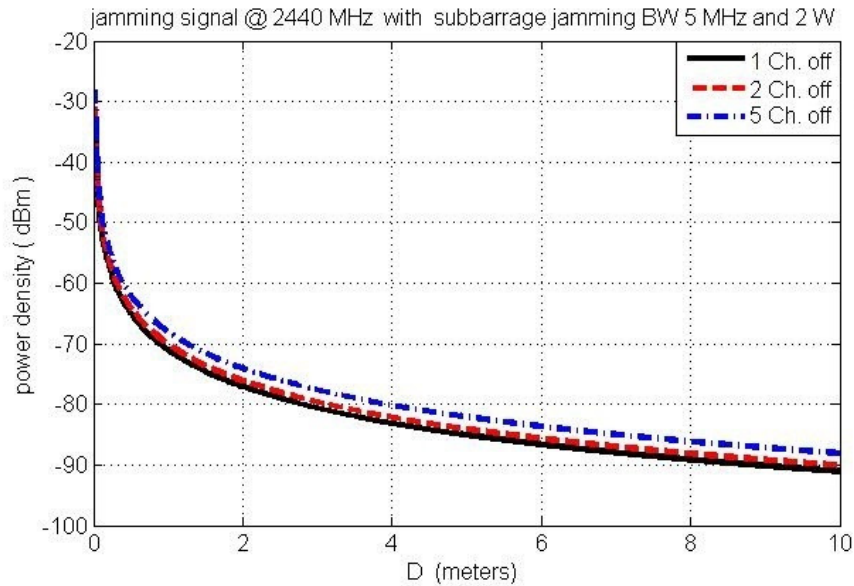

Fig 6. Jamming Signal Level Increase with Time Duty to AFH

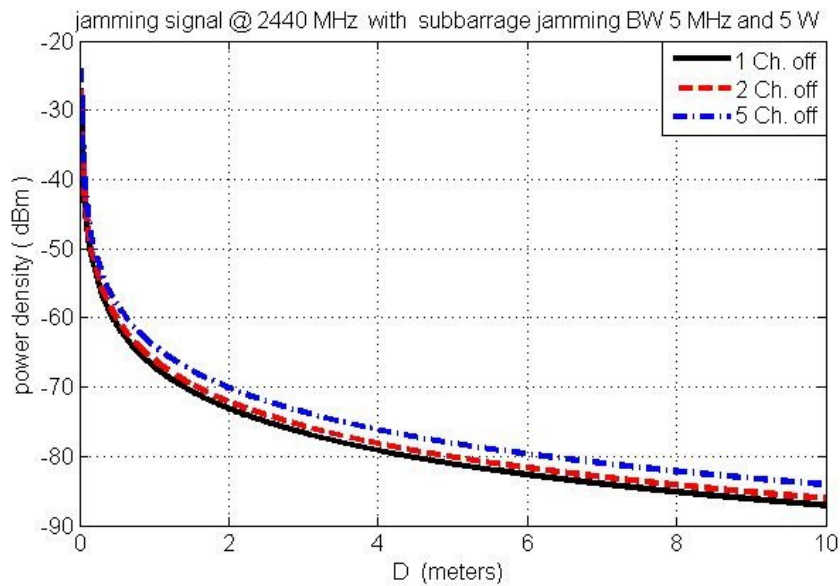

Fig 7. Jamming Signal Level Increase with Time Duty to AFH

## 5. CONCLUSIONS

The Bluetooth jammer is one of the challenges for broken the frequency hopping spread spectrum which have a processing gain (19) dB as immunity of jamming signal, and Bluetooth system has a technique is called AFH which is remove all channels interfere with anther frequencies or effected by jamming signal.

This paper proved the block jammer signal (noise powered with different bandwidth 5MHz and 20MHz) not effected to the Bluetooth network and it need high power to overcome the processing

173

gain and the path loss attenuator. The effect appears within 1 meter distance by increasing the power density of the jamming signal, the bandwidth of the jamming signal was set to 5 MHz and uses very slow swept noise to effect the channels sequentially and kept that channel out of service according to AFH, and that will reduce the processing gain with time.

According to future aspect, a new technology need to be tested such as the follower jammer technique, which is effected on frequency hopping.

**Authors**

**Adil H. M. Aldlawie**, received the B.Sc. degree in Electrical and Electronic Engineering from Iraq, Baghdad in al University of Technology in 1987, and the M.Sc. degree in Electrical and Electronic Engineering in University of Technology in 2004. His research interests include Telecommunication, Warfare System, Satellite Communications, and Electronics.

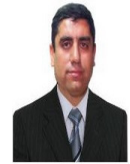

**Ghassan A. QasMarrogy**, received the B.Sc. degree in Computer and Communication Engineering from Iraq, Baghdad in al MANSOUR University College in 2009, and the M.Sc. degree in Electrical and Electronic Engineering in Eastern Mediterranean University in 2013. His research interests include Data communication and Telecommunication, Wireless Networks, MANETs, Cryptography and network security, and Multimedia Communications.

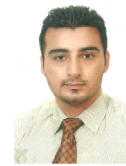